\documentclass{svjour3}                     % onecolumn (standard format)
\smartqed  % flush right qed marks, e.g. at end of proof
\usepackage{graphicx}
\usepackage{amsmath}
\usepackage{amssymb}
\usepackage{cite}
\usepackage{subfig}
\newcommand*{\pbar}{\ensuremath{\overline{\rm{p}}}}
\newcommand*{\pbHe}{\ensuremath{\overline{\rm{p}}\rm{He}^+}}
\begin{document}

\title{Preliminary Results from Recent Measurements of the Antiprotonic Helium Hyperfine Structure
}


\author{T. Pask}


\institute{T. Pask \at
              Stefan Meyer Institute for Subatomic Physics, Austrian Academy of Sciences, Boltzmanngasse 3, A-1090 Vienna, Austria \\
              Tel.: (+43 1) 4277 29725\\
              Fax: (+43 1) 4277 9297\\
              \email{thomas.pask@cern.ch}           %  \\
%             \emph{Present address:} of F. Author
}

\date{Received: date / Accepted: date}

\maketitle

\begin{abstract}
We report on preliminary results from a systematic study of the hyperfine (HF) structure of antiprotonic helium. This precise measurement which was commenced in 2006, has now been completed.  Our initial analysis shows no apparent density or power dependence and therefore the results can be averaged.  The statistical error of the observable M1 transitions is a factor of 60 smaller than that of three body quantum electrodynamic (QED) calculations, while their difference has been resolved to a precision comparable to theory (a factor of 10 better than our first measurement).  This difference is sensitive to the antiproton magnetic moment and agreement between theory and experiment would lead to an increased precision of this parameter, thus providing a test of CPT invariance.

\end{abstract}

\section{Introduction: Antiproton spin magnetic moment}\label{intro}

The most fundamental particles, from which all visible matter in the universe is comprised, are spin-half particles called fermions.  All fermions have antiparticle equivalents which have identical properties but opposite charge.  Matter and antimatter do not however exist in equal quantities.  The majority of the observable universe consists of matter while antimatter exists usually only as a result of energetic particle interactions.

Particles and antiparticles are theoretically symmetric under the following simultaneous operations called $CPT$ invariance:

\begin{itemize}
\item Charge conjugation, $C\psi(r,t) \, = \, \overline{\psi}(r,t)$
\item Parity change, $P\psi(r,t) \, = \, \psi(-r,t)$
\item Time reversal, $T\psi(r,t) \, = \, \psi(r,-t)$.
\end{itemize}

An implication of $CPT$ invariance is that all the parameters of a particle must have identical absolute values to those of its antiparticle.  Therefore an experimental measurement comparing these parameters between particles and their antiparticle equivalents, constitutes a test of this theory.

The gyromagnetic ratio of a proton $g^{\rm{p}}_{s}$ has been measured to $\sim$~10~ppb, summarised in Ref.~\cite{Codata}. In principle it is possible to measure $g^{\rm{p}}_{s}$ in a Penning trap \cite{GG,Rod,Quint} to increase this precision. The value can then be compared to that of an antiproton $g^{\rm{\pbar}}_{s}$ in the same trap.  Until this method has been proved experimentally, the only environment where the antiproton spin magnetic moment can be measured is in antiprotonic atoms.  Indeed the current most precise measurement ($(g^{\rm{p}}_{s}-g^{\pbar}_{s})/g^{\rm{p}}_{s} \, < \, 0.3\%$) was performed on antprotonic lead \cite{pb}.  In 1996, the ASACUSA collaboration measured laser transitions between the (37,35)~$\rightarrow$~(38,34) states of antiprotonic helium resolving the Hyperfine structure~\cite{Wid97} leading to a measurment of the antiproton-proton spin magnetic moment ratio to a precission of 1.6$\%$~\cite{HFS} in 2001.

Antiprotonic helium $(\pbHe)$~\cite{Iwasaki,Yam93,blue,newblue} is a neutral three body system consisting of one electron, one helium nucleus, and one antiproton (He$^{++}$ + e$^-$ + $\pbar$)~\cite{condo,russell}.  When an antiproton approaches a helium atom, at an energy of the order of the helium ionisation energy, it can, simultaneously, eject one of the two ground state electrons and become captured.  97\% of the captured antiprotons anhihilate within nanoseconds with one of the nucleons in the nucleus but $\sim$~3\% occupy metastable states with lifetimes of the order of $\sim$~3~$\mu$s.  Due to its long life time this unique particle is an ideal subject for spectroscopy.

A \emph{hyperfine (HF)} structure~\cite{blue}  arises from the interactions between antiproton spin, its angular momentum and electron spin.  The magnetic moment of the antiproton can be determined by the comparison of the measured HF transition frequencies with three-body QED calculations.  As a result such a measurement constitutes a rigorous test of the theory.

\section{Antiprotonic helium hyperfine structure}\label{sec:1}

\begin{figure}[h]
\includegraphics[scale=0.6]{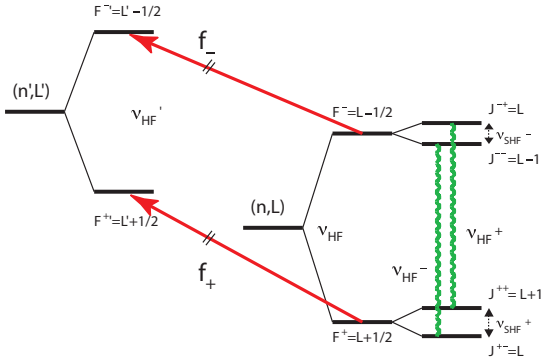}
% ----------------------------------------------------------------
\caption{Schematic view of the splitting of $\pbHe$ for the unfavoured electric dipole transitions.  The state drawn on the right is the radiative decay dominated parent $(n,L)$, and the left state is the Auger decay dominated daughter $(n',L')$. The laser transitions, from the parent to daughter doublets, are indicated by straight lines and the microwave transitions, between the quadruplets of the parent, by wavy ones.  For this experiment $(n,L)$~=~(37,35) and $(n',L')$~=~(38,34)~\cite{pask}.}
\label{fig:trans}
\end{figure}

The HF splitting (Fig.~\ref{fig:trans}) of $\pbHe$ has been calculated by Korobov and Bakalov \cite{BK,KB,K,YK} to the $\alpha^{4}$ order.  The dominant HF splitting is caused by the interaction of the antiproton orbital angular momentum $\vec{L}$ with the electron spin  $\vec{S}_e$ ($\pbar$~-~e$^{-}$ spin-orbit splitting). This is referred to as a \emph{hyperfine} rather than \emph{fine} splitting because the perturbation is caused by the interaction of two different particles.  A further \emph{superhyperfine (SHF)} splitting, shown in Fig.~\ref{fig:trans}, is caused by the combination of three interaction: $\pbar$ spin-orbit splitting (coupling of the antiproton orbital angular momentum and the antiproton spin $\vec{S}_{\pbar}$), the contact spin-spin and tensor spin-spin interactions.

% ----------------------------------------------------------------

The HF doublet is described by the quantum number $\vec{F} \, = \, \vec{L} \, + \, \vec{S}_{e}$ and the quadruplet by $\vec{J} \, = \, \vec{F} \, + \, \vec{S}_{\pbar}$, shown in Fig.~\ref{fig:trans}.  Between these substates there are two M1 transitions $\nu^{+}_{\rm{HF}}$ and $\nu^{-}_{\rm{HF}}$ which cause an electron spin flip and can be induced by an oscillating magnetic field:

\begin{subequations}
\begin{equation}
\nu^{+}_{\rm{HF}}: \, J_{++} \, = \, F_{+} \, + \, \frac{1}{2} \, \leftrightarrow \, J_{-+} \, = \, F_{-} \, + \, \frac{1}{2},
\end{equation}

\begin{equation}
\nu^{-}_{\rm{HF}}: \, J_{+-} \, = \, F_{+} \, - \, \frac{1}{2} \, \leftrightarrow \, J_{--} \, = \, F_{-} \, - \, \frac{1}{2},
\end{equation}
\end{subequations}

The difference between the two transitions $\Delta \nu_{\rm{HF}}$~=~$\nu^{-}_{\rm{HF}}$~$-$~$\nu^{+}_{\rm{HF}}$, is dependent on the antiproton magnetic moment and therefore it is the comparison between experiment and theory of this value that provides a test of \emph{CPT} invariance.

Bakalov and Widmann calculated the sensitivity by which individual states depend on the $\pbar$ spin magnetic moment and therefore the optimal candidates for measurement~\cite{BW}.  Some of these states are not practical due to limitations in laser capability.  Others, like the $(n,l)$~=~(39,~35) state, which are within laser capabilities, have favoured transitions to the daughter state and therefore a very narrow HF splitting ($\Delta f_{\rm{HF}}$~=~0.5~GHz), even in comparison to current laser line widths.  The previously measured (37,~35) state remains the best candidate for a precision study because there is an easily stimulated unfavoured laser transition between the $(n,l)$~=~(37,~35) and (38,~34) states.  The precision of the predicted splitting for this transition, and therefore the required experimental precision, is 33~kHz.

\section{Laser-microwave-laser spectroscopy} \label{sec:2}

\noindent The experimental method~\cite{pask} is a three step laser-microwave-laser process where

\begin{itemize}
\item a population asymmetry is induced by depopulating the $F^+$ doublet with resonant  laser light, shown as straight lines in Fig.~\ref{fig:trans}.
\item A microwave pulse transfers the population from the $F^-$ to the $F^+$ doublet, if it is resonant with either the $\nu^+_{\rm{HF}}$ or $\nu^-_{\rm{HF}}$ transitions, shown as wavy lines in Fig.~\ref{fig:trans}.
\item Then a second laser, tuned to the same frequency as the first, measures the new $F^+$ population.
\end{itemize}

The experiment was carried out at the Antiproton Decelerator (AD) at CERN, which provided a beam of $1-4 \, \times \, 10^7$ antiprotons with a pulse length of 200~ns (FWHM) and kinetic energy 5.3~MeV.  The target gas was contained in a cylindrical microwave cavity surrounded by a cryostatic gas chamber that could be cooled to a temperature of $\sim$~6~K.  Two Cherenkov counters covered 2$\pi$ steradians around the target from which the signal was amplified by fine-mesh photomultipliers (PMTs)~\cite{PMT}.

A continuous wave (cw) laser beam of wave-length 726.1~nm was split into two seed beams.  The two laser pulses were produced by amplifying the seed beams with dye filled Bethune cells, pumped by pulsed Nd:YAG lasers~\cite{pe}.  The second laser pulse was delayed by a time $T$ after the first.  The pump beams were stretched so that the two pulse lengths were of the same order as the Auger lifetime.

The required microwave signal was created by Port~1 of a vector network analyser (VNA), amplified with a pulsed travelling wave tube amplifier (TWTA)  and transported to the target through a rectangular waveguide~\cite{App}.  A cylindrical resonant microwave cavity contained the microwave field at the target.  To cover the entire microwave range the cavity was over coupled to the waveguide so that its resonance was broad.

Previously, different choke positions of a triple-stub-tuner (TST) were used to match the impedance of the waveguide to that of the cavity for a range of frequencies~\cite{App}.  This time, a constant microwave power $P$ was produced at the target by firing a predetermined power down the unmatched waveguide.  Most of the signal was reflected and dumped to a 50~$\mathrm{\Omega}$ terminator by a three-way circulator.  This removed standing waves from the system and allowed the relatively small amount of power absorbed by the cavity to be controlled to within 1~dB over the frequency range.

\section{Results} \label{sec:3}

Density effects produce a systematic shift of the laser transitions \cite{Torii99}.  In earlier laser spectroscopy experiments, this source of uncertainty was minimised, first by a linear extrapolation to zero density~\cite{Hori01}, then later through the use of an ultra low target density ($\sim~10^{17}$~cm$^{-3}$)~\cite{Hori03}.  The stopping distribution at low densities is larger than the cavity depth and therefore a high density target ($\sim$~$3 \times 10^{20}$~cm$^{-3}$) must be used.  Fortunately the density shift is calculated to be considerably smaller (80~kHz/250~mbar) in the microwave spectrum~\cite{kman}.  The first measurement of the HF splitting was performed at two different target pressures ($p$~$\sim$~250~mbar and $p$~$\sim$~500~mbar)~\cite{HFS,HFShift} and had an uncertainty of $\sim$~300~kHz, far too large to observe the predicted splitting.  The most recent published result resolved $\Delta \nu_{\rm{HF}}$ to $\sim$~60~kHz~\cite{pask} but, although many other systematic effects were studied~\cite{Pbar08}, was determined from measurements made at only one target pressure ($p$~$\sim$~250~mbar).  Another possible source of uncertainty was the microwave power, although no significant shift has been predicted.  The new measurments address these systematics, in particular density and microwave power effects.

% ----------------------------------------------------------------
\begin{figure}[h]
\subfloat[]{
\label{fig:subfig:scan_p}
\includegraphics[scale=0.3]{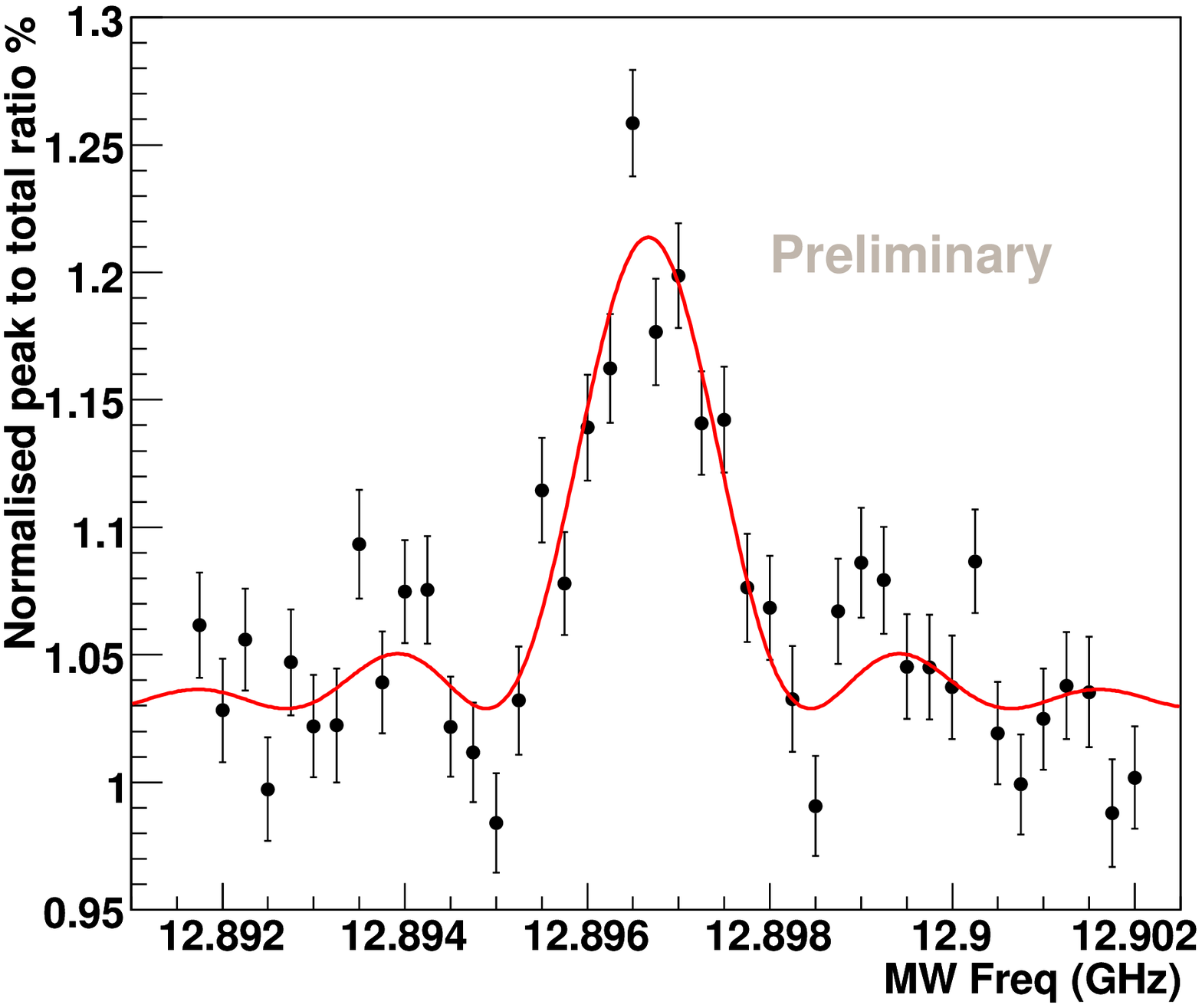}}
\subfloat[]{
\label{fig:subfig:scan_n}
\includegraphics[scale=0.3]{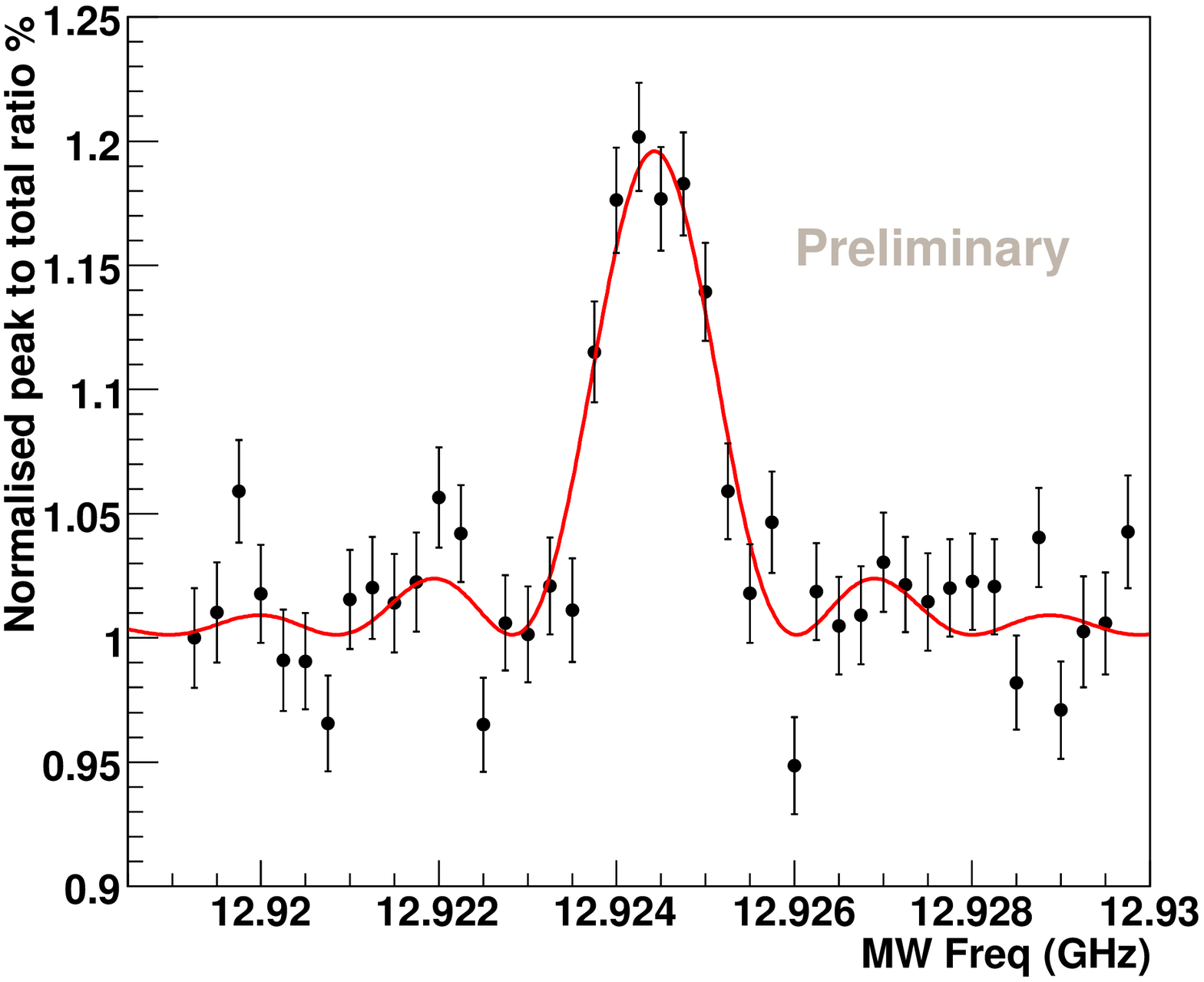}}
% ----------------------------------------------------------------
\caption{Averaged data from resonant profiles measured at $p$~=~150~mbar, $T$~=~500~ns and $P$~$\sim$~5~W. (a) The $\nu^{+}_{\rm{HF}}$ resonant transition. (b) The $\nu^{-}_{\rm{HF}}$ resonant transition.}
\label{fig:scan}
\end{figure}

Microwave frequency profiles were measured at $p$~$\sim$~150~mbar and $\sim$~500~mbar at $T$~=~350~ns.  At $p$~$\sim$~150~mbar, additional profiles were measured at $T$~=~500~ns and 200~ns.  Figure~\ref{fig:scan} shows the averaged data for the highest statistics measurements.

\begin{figure}[h]
\includegraphics[scale=0.3]{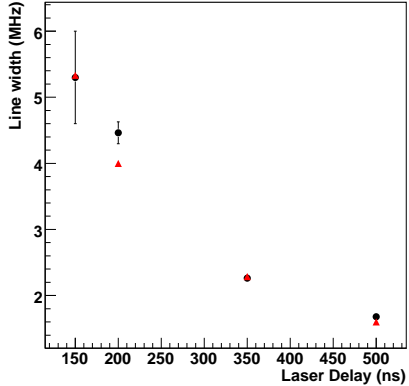}
% ----------------------------------------------------------------
\caption{The linewidth as a function of $T$, where the FWHM of measured frequency scans are shown as circles and the widths calculated from the Fourier limit, FWHM~=~0.799/$T$, as triangles. The data were measured at a p~=~150~mbar, 250~mbar and 500~mbar at a temperature of 6.1 K.}
\label{fig:width}
\end{figure}

Collisional and power broadening have been shown to have little effect on the line width, the dominating effect is the Fourier transform of the microwave pulse length which is equivalent to the laser delay $T$, see Fig.~\ref{fig:width}.  Analysis of the new results is not yet complete but preliminary observations show that a density dependent shift of the transition frequencies is much smaller than predicted and still unresolvable despite a factor of 10 increase in the precission.  There was also no observable power dependence.

% ----------------------------------------------------------------

This preliminary analysis indicates that, although the difference between theory and experiment $\delta_{\rm{th-exp}}$ for the $\nu^{\pm}_{\rm{HF}}$ lines is of the order $\sim$~300~kHz, it is  a factor of three smaller than the theoretical error~\cite{KB, BW}.  It should be noted that the experimental errors are expected to be $\sim$~20~kHz and therefore a factor of 60 smaller than the theoretical ones.  The difference between the lines $\Delta \nu_{\rm{HF}}$ seems to be in good agreement with theory and, by averaging the results, an error with a similar statistical precision as the theoretical value can be achieved.  After further analysis, however, the error bars could still be increased by currently unidentified systematics.  The theoreticians are working on a calculation to the $\alpha^{6}$ order.

\begin{acknowledgements}
The Author would like to thank the two undergraduate students: Peter Somkuti and Katharina Umlaub who worked on this project.  We are grateful to the AD operators for providing the antiproton beam.  This work was supported by Monbukagakusho (grant no. 15002005), by the Hungarian National Research Foundation (NK67974 and K72172) and by the Austrian Federal Ministry of Science and Research.
\end{acknowledgements}

\end{document}